\documentstyle[aps,prl,psfig]{revtex}
\begin{document}
\topmargin=0.0cm

\twocolumn[\hsize\textwidth\columnwidth\hsize\csname@twocolumnfalse\endcsname

\title{Polyethylene under tensile load: 
strain energy storage and breaking
of linear and knotted alkanes 
probed by first-principles molecular dynamics calculations}

\author{A. Marco Saitta~\cite{e-mail1} and Michael L. Klein }

\address{Center for Molecular Modeling, Dept. of Chemistry,
University of Pennsylvania, Philadelphia, PA 19104-6202}

\date{\today}
\maketitle

\begin{abstract}
The mechanical resistance of a polyethylene strand 
subject to tension and the way its
properties are affected by the presence of a knot
is studied using first-principles molecular dynamics calculations.
The distribution of strain energy for
the knotted chains has a well-defined shape
that is very different from the one found in the linear case.
The presence of a knot significantly weakens the chain
in which it is tied. Chain rupture invariably 
occurs just outside the entrance to the knot,
as is the case for a macroscopic rope.

\end{abstract}

\pacs{PACS numbers: 
}
]
\narrowtext

\section{Introduction}

Structural isomerism is a well-studied phenomenon 
in chemistry, but it is also possible to construct isomers of molecules 
that differ not in their connectivity, but in their topology
\cite{frisch-wass,schill,walba}. 
These topological isomers exhibit different optical rotary power.
Examples of topologically complicated structures are knots,
for which the simplest is the ``trefoil''. 
A trefoil is constructed from one
chain that forms a single, self-threaded loop~\cite{ashley,degennes}.
Technically, the ends of the chain must be connected for this
structure to be a knot~\cite{knots}.
DNA fragments, for example, have been observed to form knots similar to
the trefoil as well as more complicated ones~\cite{dna}. Interpenetrating
entangled polymers also form knots. Statistical arguments based on
self-avoiding random walks, Hamilton cycles, and so on, have
demonstrated conclusively that the chain length is the crucial factor
governing the probability of knot formation
\cite{russia-nature,michels}. As the chain length
increases, this probability goes to one exponentially
\cite{whittington}.

The self-entanglement of polymer chains has profound effects on the
physical properties of the bulk material, such as its crystallization
behavior, elasticity, and rheology. 
For example, the transfer of a knot from a polymer in the
melt state to that in the crystalline/glassy state has been studied
\cite{bayer}. Crystallization of polymers is inhibited 
by the presence of knots and the freezing tends to remove them. On the other
hand, quenching the melt to a glass tightens and stabilizes existing knots.
The structure and stability of knots in polymers has been a subject of
extensive study~\cite{iwata,mansfield,mans2,frisch-new}.
We recently reported a first-principles molecular dynamics
study of a trefoil knot in a polyethylene strand, which revealed
a surprising similarity in behavior between 
macroscopic and microscopic knotted strands under tensile
stress~\cite{noi}.
In the present work, we focus on the storage of strain energy
in a polymer.
In particular, we will describe how strain distributions differ
between linear and knotted polymer chains.
Further, this work demonstrates that an efficient combination
of classical~\cite{goddard} and density-functional based first-principles
\cite{cp} molecular dynamics (MD) can be a very
useful tool in the description of physical as well as chemical
properties of polymers with and without topological defects.

The essential details of the calculations are provided in the following
section. The results for the linear polyethylene-like (alkane) molecules are
reported in Section~\ref{linear}, while classical and first-principles
simulations on the knotted strands are presented in Section~\ref{knot}.
We discuss our findings in Section~\ref{concl}.

\section{Computational method}\label{compmeth}

The first principles calculations were performed using the
Car-Parrinello (CP) method~\cite{cp}, which
combines MD and density-functional theory (DFT).
DFT is known to provide an accurate and reliable
description of carbon- and hydrocarbon-based systems~\cite{sandrolo}.
We adopted Becke and Lee-Yang-Parr (BLYP)~\cite{blyp} gradient corrections
to describe the exchange and correlation functionals, respectively.
The electron-ion interaction was described by pseudopotentials
in the Martins-Troullier~\cite{mt} form. The plane-wave basis set
had a cutoff of 60 Ry, with a $\Gamma$-point integration
in the Brillouin zone.
Electron spin was explicitly taken into account~\cite{lsda},
to obtain more reliable values of the dissociation energies.
We described the isolated molecules with
periodic boundary conditions and supercells that were large enough to
avoid any significant interaction with periodic images.
The same, large, tetragonal supercell ($12~\AA\times 12~\AA\times 20~\AA$)
was used in both the linear and knotted cases.
While these dimensions may be oversized for the linear molecules, 
this approach afforded us consistent and comparable results.

Classical MD simulations were carried out according to the united
atom (UA) scheme~\cite{CJM}, in which every methylene ($-CH_2-$) or
methyl ($-CH_3$) group is considered a ``pseudoatom''. The potential
of interaction
included harmonic bond-stretching, harmonic bond-bending, 
a torsional field, and a Lennard-Jones
interaction between non-bonded atoms. We fitted the stretching and
bending interaction constants of the model to {\it ad hoc}
CP calculations on small alkanes.

In both classical and first-principles MD simulations, 
external tensile stress was mimicked
by constraining the chain end-to-end distance $L$ to a fixed value.
The system was evolved dynamically at room
temperature before being annealed in order to find the (constrained)
minimum. 

\section{Linear alkanes}\label{linear}

All the results reported in this section refer to CPMD simulations.
As a first step, we studied a decane molecule ($C_{10}H_{22}$),
which can be taken as representative of longer linear alkanes
and, by extrapolation, of polyethylene.
The equilibrium
structural parameters, reported in Table 1, are in excellent
agreement (within 1\%) with other DFT
results and with experimental data~\cite{colino1,colino2} for
hydrocarbons and/or crystalline polyethylene (PE).

\begin{figure}
\centerline{\psfig{figure=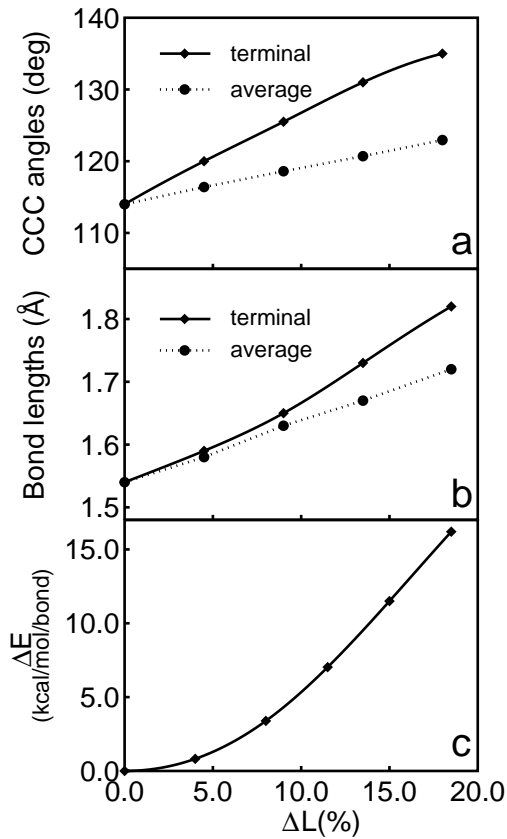,width=8.25truecm}}
\caption{Analysis of the strain distribution along the $n$-decane
molecule as function of the linear deformation. In panel {\bf a}
shown is the bond angle (degrees) of the terminal carbons
(diamonds, solid line), and the average {\em other} $C$-$C$-$C$ bond
angles
(circles, dotted line). In panel {\bf b} we report the $C$-$C$
bond-length ($\AA$) distribution accordingly.
In panel {\bf c} the average strain energy (kcal/mol)
per $C$-$C$ bond is displayed.
}
\label{fig1}
\end{figure}

The molecule was stretched in a series of steps.
Additional external load was applied to the system
by successively increasing $L$. To do so, the nuclear
coordinates were all scaled uniformly
along the molecular $z$ axis.
In the first stages of tensile loading, the geometry of the system
was optimized by minimizing the forces acting on atoms, but no
dynamics was used. Quite unexpectedly, the strain did not distribute
uniformly along the chain, but tended to concentrate equally on the
two extremal bonds, where the force originating from the constraint was
actually applied. 
This behavior can be clearly observed by comparing the terminal
$\widehat{CCC}$ bond angles and $C$-$C$ bond lengths with the average
values calculated for atoms in the interior of the chain. These results
are reported in panels {\bf a} and {\bf b} of Fig.~\ref{fig1}
as a function of the distance $L$. 
For deformations $\Delta L$ larger than 10\% of the
equilibrium length $L_0$, we followed the procedure, described in the
previous section, of letting the molecule evolve dynamically at room
temperature, and then cooling it down with a simulated annealing technique.
The global constrained minimum of the system, after the CPMD evolution, 
was found by geometry optimization of the structure.
The linear $n$-decane does not break for distortions 
up to $\Delta L \sim$ 18.0\%. 

\begin{table}
\begin{tabular}{|l|c|c|c|c|}
  & ${\rm C_{10}H_{22}}$
  & ${\rm C_{10}H_{22}}$
  & Cryst. PE
  & Cryst. PE \\
  & (BLYP)
  & (BP)
  & (BP)
  & (exp.) \\
\hline
 $d_{\rm CC} $ & 1.54 & 1.53 & 1.54 & 1.55(2) \\
\hline
 $d_{\rm CH} $ & 1.10 & 1.10 & 1.11 & 1.09(1) \\
\hline
 $\alpha_{CCC}$ & 114 & 113 & 114 & 111(3) \\
\hline
 $\alpha_{HCH}$ & 107 & 108 & 106 & 108(2) \\
\end{tabular}
\caption{
Equilibrium structural parameters of $n$-decane
as determined in the present work (DFT-BLYP)
or using a Becke-Perdew (DFT-BP) functional for the
exchange and correlation (23), compared to theoretical (DFT-BP) and
experimental results for crystalline polyethylene (PE) (23,24).
}
\label{struct}
\end{table}

The distribution of strain energy of the system was estimated 
by adopting the UA model~\cite{CJM} to analyze the configurations 
provided by the first-principles calculations.
In the present case, contributions to the
total strain energy originating from the torsional potential
and the long-range non-bonded interactions are negligible.
Bond stretching and bending largely dominate in the storage
of tensile load in linear molecules.
The system can sustain about 16.2 kcal/mol per
$C$-$C$ bond before rupture (Fig.~\ref{fig1}, panel {\bf c}).

\begin{figure}
\centerline{\psfig{figure=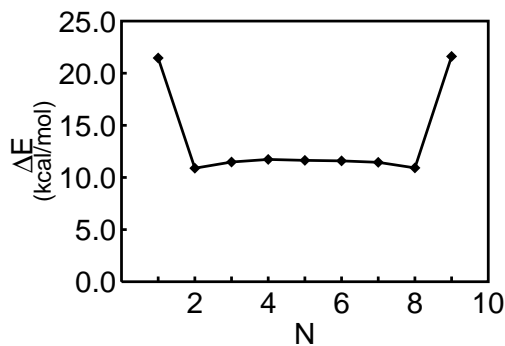,width=8.25truecm}}
\caption{Strain energy distribution along the $C$-$C$ bonds of the
$n$-decane molecule for a deformation $\Delta L=18\%$,
estimated by making use of the UA model (see text). The stress
is seen to concentrate most on the extremal bonds, while being almost
constant along the others.}
\label{fig2}
\end{figure}

The analysis of the strain energy distribution along the chain
for $L=1.18~L_0$, shown in Fig.~\ref{fig2}, confirmed the results
previously discussed, {\it i.e.}, the stress was not uniform along the
molecule. Instead, it was mostly concentrated on the two extremal bonds,
which store 20-25 kcal/mol of the energy associated with the
distortions from the ideal geometry.
All the other ``central'' bonds sustain a lower stress, 
whose average energy is about 12 kcal/mol. 

\begin{figure}
\centerline{\psfig{figure=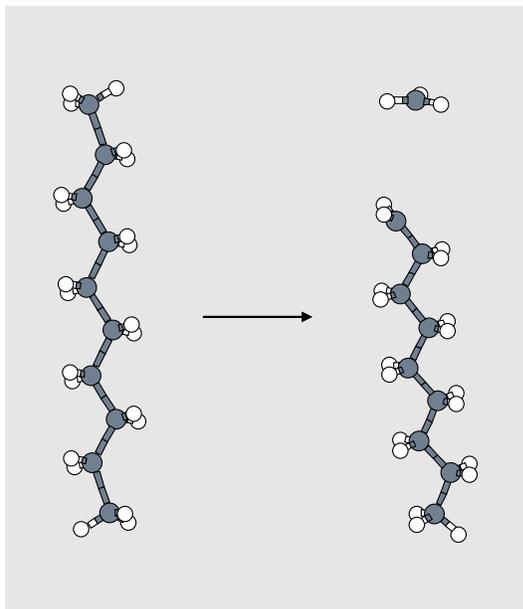,width=7.0truecm}}
\caption{A strained ($L=1.18~L_0$) $n$-decane molecule
before (left) and after (right) bond breakage.
Carbons are displayed in dark gray and hydrogens in
white. The $sp^3 \rightarrow sp^2$ change of the hybridization status of
the
atoms involved is detectable from the coplanarity
of the two carbons with their bonds.}
\label{fig3}
\end{figure}

The qualitative features of the
strain energy distribution were essentially the same for any value
of length deformation, $\Delta L$. The bonds closest to the two extremal ones
typically were about 6\% energetically
less strained than the average of the central
bonds. Thus, from the third atom on, the system can be considered
to be ``bulk'', and the strain practically constant.
Additional calculations performed on a linear $C_{21}H_{44}$ molecule indicate
that the shape of the strain energy distribution reported in
Fig.~\ref{fig2} for $C_{10}H_{22}$ is unaffected by the
size of the system. The amount of strain energy on the extremal and the
``bulk'' bonds are within the errors of the values obtained for
the shorter alkane.

As foreshadowed by the classical strain energy distribution, the CPMD
simulation results demonstrated that, for larger elongations,
the alkane molecule actually breaks at one of the extremal bonds.
As shown in Fig.~\ref{fig3}, the $C_9H_{19}{\bf \cdot}$
radical contracts very quickly after the break. 
While large deviations from equilibrium
bond lengths were found throughout the simulations, the true marker for
bond dissociation was the change in hybridization
($sp^3 \rightarrow sp^2$) for the
previously bonded carbon atoms.
Indeed, as shown in the right side of Fig.~\ref{fig3},
the three bonds about those carbon atoms become coplanar after
the break.
The length of the bond formed between the last carbon
atom of the $C_9H_{19}{\bf \cdot}$ radical 
and its closest neighbor decreases as well,
oscillating around the typical $C_{(sp^3)}$-like value,
{\it ca.}~1.49 \AA.
The experimental enthalpy of dissociation of a methyl group 
from smaller $n$-alkanes is {\it ca.} 86 kcal/mol.
Our theoretical dissociation energy of 83 kcal/mol, calculated as the energy
difference between the optimized $C_{10}H_{22}$ molecule and
the optimized radicals, gives a very good
quantitative description of the dissociation phenomenon.

\begin{figure}
\centerline{\psfig{figure=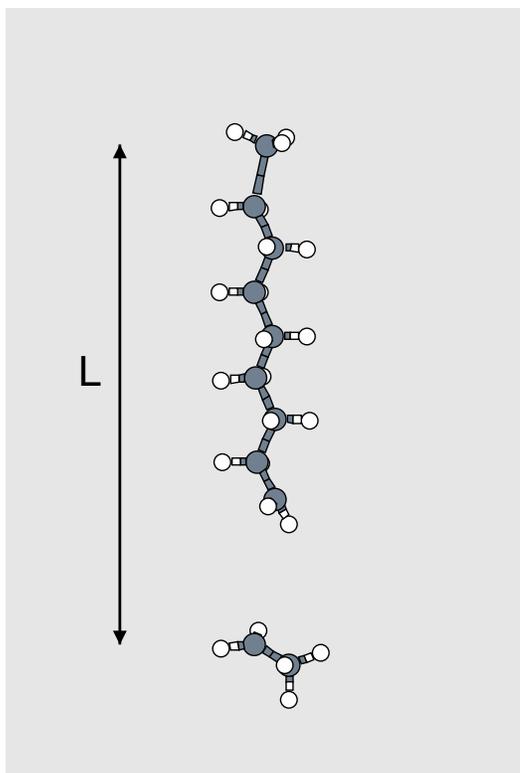,width=7.0truecm}}
\caption{The $n$-undecane molecule after bond breakage. As displayed,
the
constraint on the distance $L$ is applied between $C_1$ and $C_{10}$.
The resulting $sp^2$ hybridization of $C_9$ and $C_{10}$
is clearly observable.}
\label{fig4}
\end{figure}

The finding that a
linear alkane subject to tensile stress breaks where the external force
is applied was checked by performing analogous simulations
with a $n$-undecane molecule ($C_{11}H_{24}$). In this case, however,
the constrained distance $L$ was not the end-to-end length of the chain,
but still the $C_1$-$C_{10}$ distance. Indeed, this system behaved very
similarly. As was observed for the $n$-decane molecule, 
the chain once again ruptured where the constraint was applied. 
In contrast to the previous case,
however, the simulation consistently formed an ethyl
radical ($C_{11}H_{24} \rightarrow C_9H_{19}{\bf\cdot}+C_2H_5{\bf\cdot}$)
(see Fig.~\ref{fig4}).
The present DFT value for the dissociation energy is 81 kcal/mol,
which is again in excellent agreement with the
experimental dissociation enthalpy of an ethyl group
which is about 82 kcal/mol.
We note here that similar
calculations that do not take into account spin variables give
the same qualitative results, but the corresponding
dissociation energies (102 and 94 kcal/mol, respectively) 
are overestimated by 10-15 \%.

\section{Knotted molecules}\label{knot}

\subsection{Classical MD calculations}\label{knotMD}

The study of a knotted polymer requires
a non-trivial procedure to set up the
starting nuclear coordinates for a trefoil.
Ideally, the behavior of the system should not be biased in any way by
a specific strain distribution due to a particular choice of the initial
configuration.
Further, the computational demands of CPMD calculations
prohibit the study of loose knots because of the large system size
required.
Moreover, the initial stress must be close to, yet below, the breaking
threshold to allow the rupture to occur within
the typical time scales ($< 10~ps$) of CPMD.
With these constraints in mind,
we first performed classical MD simulations with the UA model.

\begin{figure}
\centerline{\psfig{figure=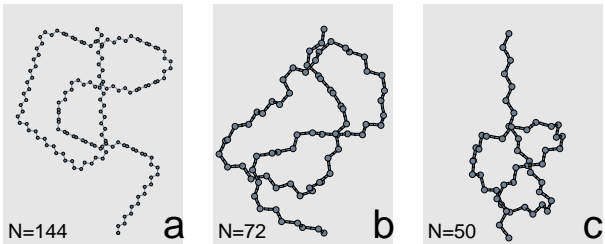,width=8.25truecm}}
\caption{Initial configurations of the chains containing
$N$=144 ({\bf a}), 72 ({\bf b}), and 50 ({\bf c}) pseudoatoms,
respectively.
}
\label{fig5}
\end{figure}

We chose as the initial system a chain of 144 pseudoatoms, and generated
the trefoil knot by adding an appropriate set of {\it gauche} defects to a
linear polymer (Fig.~\ref{fig5}, panel {\bf a}).
As we mentioned in section~\ref{compmeth}, we performed the MD
calculations by anchoring the end atoms of the polymer at a fixed 
separation distance $L$, which was then successively increased 
from one set of simulations to the next.
As the knot was tightened, fewer atoms
were directly involved in the formation of the knot and hence they
could be removed (Fig.~\ref{fig5}, panels {\bf b} and {\bf c}). 

\begin{figure}
\centerline{\psfig{figure=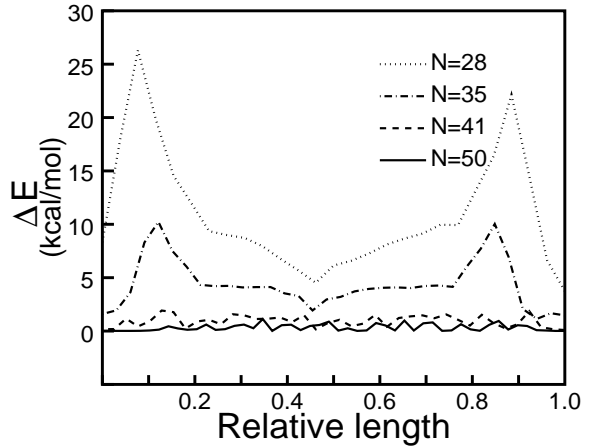,width=8.25truecm}}
\caption{Classical strain energy distribution along the chains as
obtained
after room temperature MD evolution and annealing.
Shown are results for chains containing $N$=50 (solid line), 41 (dashed
line),
35 (dotdashed line), and 28 (dotted line) pseudoatoms.
Polymer lengths (horizontal axis) are scaled in order to facilitate
direct comparison among different length molecules.}
\label{fig6}
\end{figure}

Longer simulations were performed
when the size of the molecule was reduced to $N=50$ pseudoatoms.
First, the system was kept at room temperature for at least 200 $ps$ before
cooling it down, with a slow annealing (about 50 $ps$), to $T=0~K$.
The strain-energy distribution
of each chain at the constrained (knotted) minimum energy configurations
was systematically determined for each value of the external load.
The results obtained from the CP calculations on the linear hydrocarbons
suggest that such systems can sustain a strain-energy per bond in the
range of 15-30 kcal/mol before rupture.
It is thus noteworthy that the per bond strain energy for $N=50$
chain is negligible on this ``relevant'' energy scale (Fig.~\ref{fig6}),
even though the shape of the trefoil 
is very well defined (Fig.~\ref{fig5}, panel {\bf c}).

Fig.~\ref{fig6} shows that decreasing the number of atoms to $N=41$
does not produce any noticeable effect; when $N=35$, however, the presence of
the knot begins to affect the strain distribution 
of the system in a significant way~\cite{degennes}. 
As we will see later, that strain distribution is a general fingerprint
of the presence of a trefoil; distortions from ideal geometry
tend to concentrate on the entrance and exit points of the knot,
while its central portion is much less stressed.
By decreasing of the number of pseudoatoms to $N=28$,
the system becomes very close to the ``critical'' limit. The ideal chain length
for our CPMD study thus ranges between 28 and 30 carbon atoms.
For both of these values of $N$, we performed several 500 $ps$
length classical simulations at different temperatures, which were
started from varying initial conditions. This was followed by
100 $ps$ of slow simulated annealing to obtain a
statistically meaningful set of constrained equilibrium configurations
for the trefoil under tension.
We observed that, for any given length $L$, 
deviations in the final configurations obtained from different simulations were
so small that the strain energy distribution curves could actually be
superimposed on each other with no noticeable difference.
The initial configurations for the CP simulations
were generated by averaging this set of constrained ground-state positions.

\vskip -0.5cm
\subsection{CPMD calculations}\label{knotCP}
\vskip -0.5cm

\begin{figure}
\centerline{\psfig{figure=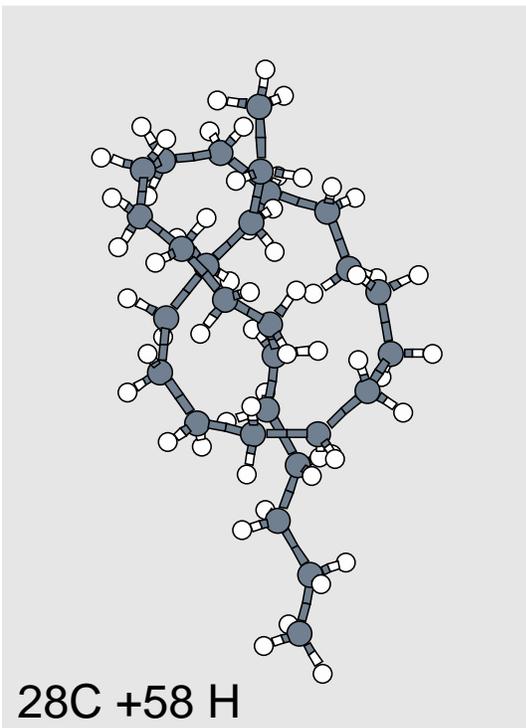,width=7.0truecm}}
\caption{
Sample of initial configuration used in the {\it ab initio}
simulations of the $C_{28}H_{58}$ molecules.
The positions of the carbon atoms are obtained from the
classical MD simulations, and hydrogens are added to each
carbon according to the appropriate tetrahedral symmetry.
}
\label{fig7}
\end{figure}

The knotted minimum energy configuration obtained from the previous
section was decorated with hydrogens atoms
according to the ideal tetrahedral symmetry
of carbon atoms in alkanes. Structural relaxation
showed that this initial guess for the hydrogen positions was very close
to those found with minimum energy geometry in the distorted system.
In Fig.~\ref{fig7} we show a sample starting configuration
of a $C_{28}H_{58}$ $n$-alkane.

CP simulations were carried out with the same
technique adopted for the linear alkanes, in which external
tensile stress was applied by constraining the end-to-end distance $L$.
Results for the $C_{28}H_{58}$ and $C_{30}H_{62}$
molecules were very similar.
Henceforth, unless specified, we will only refer to
calculations for the $N=28$ carbon alkane.
Several sets of simulations were carried out for $L=10.75,
11.50, 12.50, 13.00, 13.50, 13.75$, and $14.00~\AA$,
respectively.
For each value of $L$, the
initial positions were optimized (with the constraint)
before heating the system to room temperature. We employed
smaller timesteps and longer simulations as compared to the
linear molecules case. As explained above, this should prevent
the initial conditions from biasing the behavior of the system.
The sound velocity of polyethylene along the chain direction in the crystal
\cite{goddard} is about 18 $km/s$ and thus an estimate of the time
needed for the slowest longitudinal phonon to travel along
a 30-atom chain is about 0.25 $ps$; this is the lower bound
for the time scale of our calculations. We actually performed
CPMD simulations at least 3 or 4 times longer than this value.

\begin{figure}
\centerline{\psfig{figure=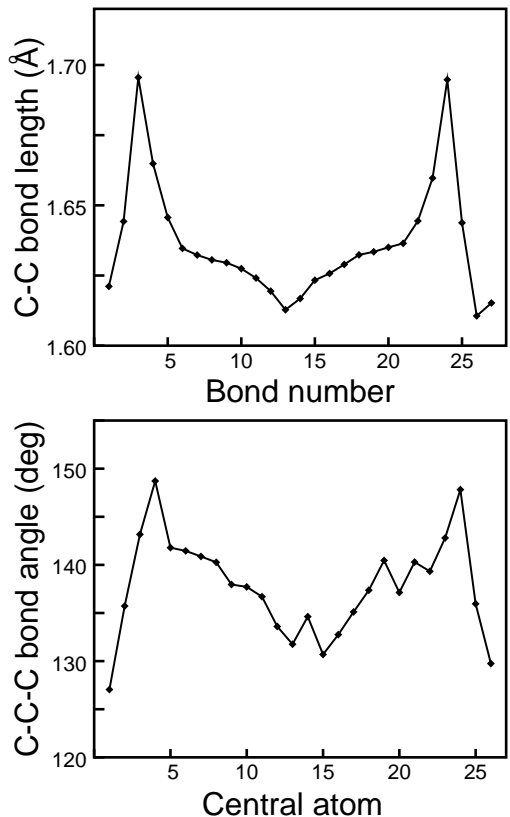,width=7.0truecm}}
\caption{
Distribution of bond lengths and angles in the $C_{28}H_{58}$
alkane at $L=13.0~\AA$. The largest deviations from the
equilibrium values occur at the entrance and exit points of
the knot.
}
\label{fig8}
\end{figure}

The optimized {\it ab initio} structure is typically more favorable by 100-120
kcal/mol as compared to the classical ground-state configuration.
The average classical strain energy distributions obtained during
finite temperature CPMD runs show a very similar shape to the classical
ground-state distributions, but with a larger amount of strain located on
the bonds at the entrance and exit points of the knot, along with a relaxation
of the central portion of the knot. This is analogous to the results reported
elsewhere~\cite{noi} and suggests that localized spikes in the strain
distribution are necessary to allow global relaxation for the remaining
portions of the carbon backbone. In analogy with the linear case, it is
interesting to note that the second bond at each end
is the least distorted of the whole chain.
This bond is neither inside the knot
nor the one at which the constraint force is applied.

\begin{figure}
\centerline{\psfig{figure=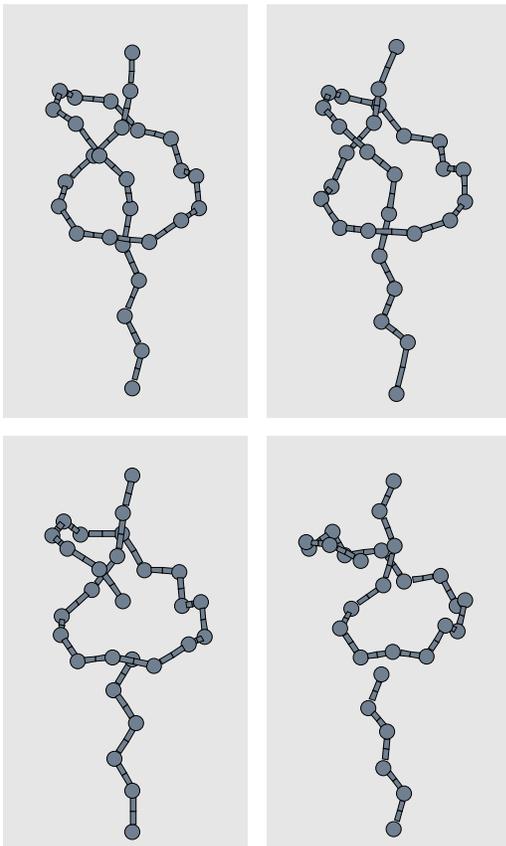,width=7.0truecm}}
\caption{
Snapshots of CP dynamical evolution of the knotted $C_{28}H_{58}$
alkane. For the sake of clarity we only display the carbon
backbone. The end-to-end distance in the simulations presently
shown is $L=13.5~\AA$, and the temperature is $T=300~K$.
}
\label{fig9}
\end{figure}

The system does not cross the dissociation barrier
at room temperature within the time scale of the simulations
for elongations smaller than $L=13.5~\AA$.
Larger and larger fluctuations of bond lengths,
up to 30 \% of the equilibrium values ($1.54~\AA$),
are observed along the dynamical evolution as $L$ increased.
This is particularly true for the bonds located at the
entrance and exit points of the knot; however, no bonds
dissociated.
In Fig.~\ref{fig8} we report an example of the $C$-$C$ bond length
and $\widehat{CCC}$ angle distributions along the molecule for $L=13.0~\AA$.
The largest deviations from equilibrium values were about 10~\%
in the lengths and 30~\% in the angles; they occurred near the
entrance and exit points of the knot. In contrast, the average deviations
in the central portion of the knot were {\it ca.}
5~\% and 18~\%, respectively.

Our data analysis indicated that monitoring the bond angles about the
two atoms forming a bond was a much more effective method to determine
chain rupture as compared to analysis of bond lengths.
Given two bonded carbon
atoms $C_i$ and $C_{j=i+1}$, we refer to the other
atoms bonded to them as $C_{i-1}, H_{i,1}, H_{i,2}$ and $C_{j+1},
H_{j,1}, H_{j,2}$, respectively.
We define $\alpha_{i,n=1,3}$ to be
the angles $\widehat{C_{i-1}C_i H_{i,1}}$, $\widehat{C_{i-1}C_i H_{i,2}}$
and $\widehat{H_{i,1}C_i H_{i,2}}$, and $\beta_{j,n=1,3}$ to be
$\widehat{C_{j+1}C_j H_{j,1}}$, $\widehat{C_{j+1}C_j H_{j,2}}$ and
$\widehat{H_{j,1}C_j H_{j,2}}$, accordingly.
$\sum_n\alpha_{i,n=1,3}$ and $\sum_n\beta_{j,n=1,3}$ equal $328.4^o$ 
in perfectly tetrahedrical
($sp^3$ hybridization) geometries, and $360^o$ in planar ($sp^2$)
molecules.

\begin{figure}
\centerline{\psfig{figure=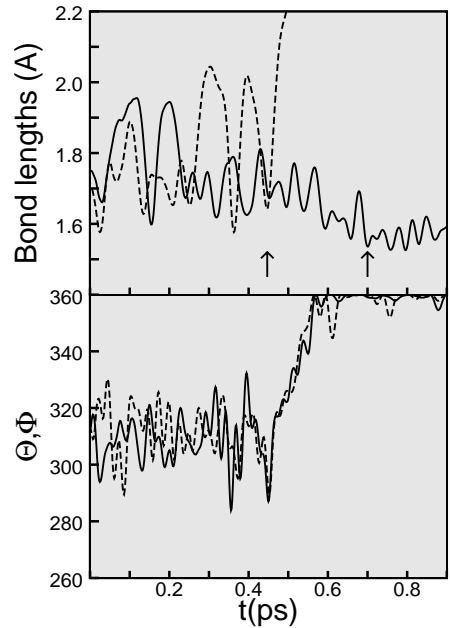,width=7.0truecm}}
\caption{
Upper panel: time evolution of the lengths of the bonds located
at the entrance and exit points of a trefoil knot with $N$=28 $C$
atoms and end-to-end distance $L=13.5~\AA$.
Lower panel: time evolution of the sums $\Phi$ and $\Theta$
of the bond angles about the atoms
involved in the rupture (see text).
Even when large bond-length deviations from equilibrium
values occurred, the chemical bond did not break
unless a change in the hybridization status of the
key atoms was observed.
Arrows indicate that the rupture and the readjustment
of the system are separated by a time interval of about
$250~fs$.
}
\label{fig10}
\end{figure}

\begin{figure}
\centerline{\psfig{figure=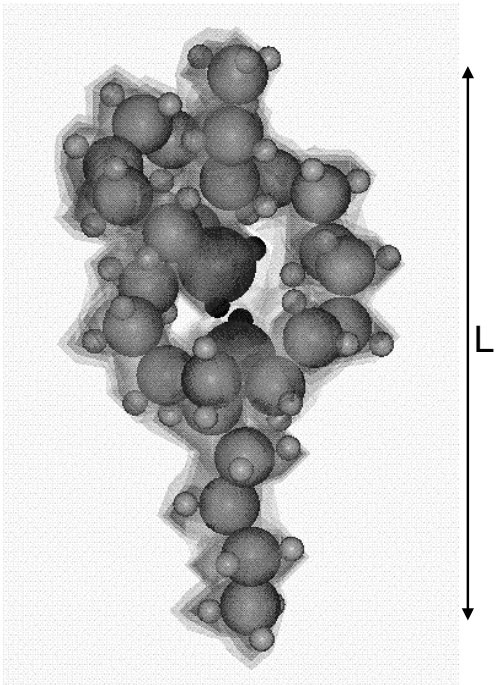,width=7.0truecm}}
\caption{
Electronic charge density of the $C_{28}H_{58}$ trefoil at the
breakpoint for $L=13.5~\AA$ and $T=300~K$. The two atoms involved
in the bond breaking are drawn with a larger size and
a darker color. A gap in the electronic density is clearly
observable.
}
\label{fig11}
\end{figure}

When $L=13.5~\AA$, as shown in Fig.~\ref{fig9}, the molecule
finally snapped during the finite temperature run
at the exit point of the knot with the two ends
then strongly recoiling back.
In the lower panel of Fig.~\ref{fig10} we report
$\Phi_{22}=\sum_n \alpha_{22,n}$ and $\Theta_{23}=\sum_n\beta_{23,n}$
as functions of time.
The fact that these quantities tended towards a value around
$360^o$ provides a simple, yet unequivocal, sign that the
hybridization status of $C_{22}$ and $C_{23}$ atoms had changed
from $sp^3$ to $sp^2$ and, thus, that their chemical bond was
broken. In the upper panel of the figure,
we display the entrance and exit bond lengths for the same
system. As previously discussed, very large deviations from
the ground-state geometry are present for {\em both} bonds;
however, only one of them finally ruptured. 
It is interesting to point out that, as marked
by the arrows, the symmetric counterpart of the breaking bond
needed about $250~fs$ to readjust and then
oscillate around the equilibrium length of $1.54~\AA$.
This is in agreement with our estimate of the typical
time scale of an alkane of this size.
An important quantity provided by first-principles MD is
the electron charge density of the system, which is particularly
useful in studies focused on the evolution of chemical bonds.
In Fig.~\ref{fig11}, a snapshot of the charge density after 0.8 ps
of dynamical evolution is reported.
Between the carbon atoms $C_{22}$ and $C_{23}$
there is a gap of charge density that confirms the
bond breaking.

Other simulations performed
using different initial conditions ({\it i.e.} heating rate,
hydrogen atom optimization, etc.), or at larger tensile stress
($L=13.75, 14.00~\AA$) gave similar results; namely,
bond breaking occurs randomly at one of the two symmetric
entrance and exit points. In the latter case, when $L=14.0~\AA$,
the molecule actually broke at {\em both} points;
the amount of external load was so high that, after the first
rupture, the system was not able to readjust quickly enough
to avoid the second break. This gave birth to three
daughter radicals rather than two.

The amount of strain energy that a trefoil knot
stores in a polyethylene chain before breaking is around
300 kcal/mol, and it has to contain at least 23 carbon
atoms to be able distribute such load along the molecule
without breaking. This {\it ab initio} estimate is in agreement
with results obtained via classical molecular dynamics~\cite{mans2}.
The average load per bond in the tightest possible knot is 
$\sim 13.3$ kcal/mol, which is about 80\% of the corresponding value 
in a linear alkane. This result, and the knot-induced location of the break
along a chain, are in good agreement with known properties
of macroscopic knots, which in turn suggests a universal behavior.

\section{Conclusions}\label{concl}
\vskip -0.5cm

In conclusion, we have presented an atomic level description of the
breaking
of a polyethylene-like strand, both with and without a knot, when
subjected
to tensile loading. Specifically, we find that the mechanical strength
and
the strain distribution in a polymer just before breaking
are
profoundly influenced by the presence of a simple knot. Moreover, a
knotted
strand, unlike the unknotted one, does not break at the point where the
tension is applied but, rather, at a point just outside the knot. Further,
the presence of a knot weakens the rope in which it is tied. Our
findings are in complete accord with the corresponding macroscopic
phenomenon exhibited by a rope, which suggests that some of the other
properties of knots known to sailors and fishermen for
centuries\cite{ashley}
may well have a counterpart in the nanoscopic world of polymers.

This work has been supported in part by National Science Foundation.
We thank E. Wasserman for focusing our attention on this problem,
and M.Parrinello and J.Hutter for fruitful discussions. 
Special thanks go to H.S. Mei for help with graphics, 
and to K. Bagchi and D. Yarne for useful suggestions.

\end{document}